\documentclass{article}

\usepackage{algorithm}
\usepackage{algorithmicx}
\usepackage{algpseudocode}
\usepackage{amsmath, amssymb}
\usepackage[english]{babel}
\usepackage{booktabs}
\usepackage{caption, subcaption}
\usepackage{csquotes}
\usepackage{graphicx}
\usepackage[final]{microtype}
\usepackage{spconf}

\title{Approximate Message Passing for\\Underdetermined Audio Source Separation}

\name{Turab Iqbal \qquad Wenwu Wang}
\address{Centre for Vision, Speech and Signal Processing, University of Surrey\\
         Email: \{t.iqbal, w.wang\}@surrey.ac.uk}

\renewcommand{\vec}[1]{\mathbf{#1}}

\DeclareMathOperator*{\argmax}{arg\!\,max}

\graphicspath{{figures/}}

\begin{document}
  \maketitle

  \begin{abstract}
  Approximate message passing (AMP) algorithms have shown great promise in sparse signal
  reconstruction due to their low computational requirements and fast convergence to an exact
  solution. Moreover, they provide a probabilistic framework that is often more intuitive than
  alternatives such as convex optimisation. In this paper, AMP is used for audio source separation
  from underdetermined instantaneous mixtures. In the time-frequency domain, it is typical to assume
  \textit{a priori} that the sources are sparse, so we solve the corresponding sparse linear inverse
  problem using AMP. We present a block-based approach that uses AMP to process multiple
  time-frequency points simultaneously. Two algorithms known as AMP and vector AMP (VAMP) are
  evaluated in particular. Results show that they are promising in terms of artefact suppression.
  \end{abstract}

  \section{Introduction}
  \label{section:intro}
  Source separation is a problem in which an unknown set of source signals must be estimated from a
  known set of mixture signals. For audio separation, this corresponds to segregating sounds
  produced by distinct entities, such as different human speakers or musical instruments. A common
  formulation of the problem is the linear instantaneous model given by
  \begin{equation}
    \label{eq:inst}
    \vec{y}(t) = \vec{A}\vec{x}(t) + \vec{w}(t),
  \end{equation}
  where $\vec{y}(t) \in \mathbb{R}^M$ denotes the mixtures, $\vec{x}(t) \in \mathbb{R}^N$ denotes
  the sources, $\vec{A} \in \mathbb{R}^{M \times N}$ is the \textit{mixing matrix} and $\vec{w}(t)
  \in \mathbb{R}^N$ is noise. The model assumes the mixtures at each instance of $t$ are
  instantaneous (memory-less), which means reverberation effects and time delays are ignored.
  Although the notation suggests the signals are in the time domain, it can also apply to signals in
  the time-frequency domain, with $t$ referring to a time-frequency point. This will be described
  in Section \ref{section:methods}.

  When $N>M$, the problem is \textit{underdetermined}, and no unique solution exists. To find a
  unique solution, the problem must be regularised by introducing a constraint, which often
  corresponds to a property of the solution known \textit{a priori}. By modelling this prior
  knowledge as a probability distribution, $p(\vec{x})$, the sources can be inferred using a
  suitable estimator. For example, the maximum a posteriori (MAP) estimate is\footnote{The
  dependence on $t$ has been omitted here and other places for simplicity.}
  \begin{equation}
    \label{eq:map}
    \vec{\hat{x}} = \argmax_{\vec{x} \in \mathbb{R}^N} p(\vec{x}|\vec{y}),
  \end{equation}
  where $p(\vec{x}|\vec{y})$ is the posterior distribution. Assuming $\vec{x}$ and $\vec{w}$ are
  independent of each other and contain i.i.d{.} components,
  \begin{equation}
    p(\vec{x}|\vec{y}) \propto \prod_{i=1}^M p(w_i-(\vec{A}\vec{x})_i) \prod_{j=1}^N p(x_j).
  \end{equation}
  The choice of the prior, $p(\vec{x})$, is important, as it affects the \mbox{separation}
  performance of the system. In the time-frequency domain, it has been shown experimentally that
  sources tend to be sparse \cite{sparse-bofill, duet-jourjine, sparse-yilmaz}, so
  choosing a prior that enforces sparsity is an effective approach. As for the choice of
  $p(\vec{w})$, we adopt the standard assumption that $p(w_i) \sim \mathcal{N}(0,
  {\gamma_\omega}^{-1})$, $i=1,...,M$, where $\gamma_\omega$ is the precision (inverse of variance)
  of the noise.

  Using this probabilistic framework, we investigate the use of approximate message passing (AMP)
  algorithms for audio separation. The original AMP algorithm \cite{amp-donoho} was proposed as an
  alternative for $L_1$ regularisation\footnote{$L_1$ regularisation corresponds to using a
  Laplacian prior.}, and was formulated as an approximation to belief propagation. It has been shown
  to converge to an exact solution when $N,M \to \infty$ and when the components of $\vec{A}$ follow
  a sub-Gaussian distribution with zero mean \cite{amp-bayati}. Its fast convergence and low
  computational complexity are reasons why it is promising. Since then, AMP algorithms have been
  developed for arbitrary priors \cite{amp-donoho1}, noise distributions \cite{gamp-rangan} and
  arbitrary $\vec{A}$ \cite{swamp-manoel, amp-rangan, ad-amp-vila, s-amp-cakmak, vamp-rangan}.

  It must be emphasised that AMP is only accurate for large $M$ and $N$, but the instantaneous
  formulation given by \eqref{eq:inst} results in small dimensions in practice. For example, a
  stereo signal containing three sources means that $M=2,N=3$. To overcome this, we propose to take
  a block-based approach in which blocks of samples are considered together rather than
  individually. This results in the dimensions of the problem being proportional to the block size.

  The paper is organised as follows. In Section \ref{section:amp}, the AMP algorithms used to
  implement audio source separation are outlined. In Section \ref{section:methods}, we describe how
  the overall system is implemented, including the time-frequency representation used and the choice
  of parameters for the AMP algorithms. In Section \ref{section:results}, the system is evaluated
  and results are presented. Finally, we summarise and suggest future work in Section
  \ref{section:conclusion}.

  \section{AMP Algorithms}
  \label{section:amp}
  Two algorithms are investigated in particular. The first is listed in Algorithm \ref{alg:amp}
  and is simply dubbed \textit{AMP}. It is based on the original algorithm, but allows an arbitrary
  prior and supports \textit{damping} \cite{amp-rangan}. The purpose of damping is to make the
  algorithm robust to instances of $\vec{A}$ that deviate from a zero-mean sub-Gaussian
  distribution, and is controlled by the damping factor, $\theta \in (0,1]$, where $\theta=1$ means
  no damping. $\vec{g}_1(\cdot)$ is a scalar function that estimates the real solution based on the
  given noisy solution, $\vec{r}_t$. For MAP estimation, its definition is given by \eqref{eq:map}.
  $\vec{g}'_1(\cdot)$ is the derivative of $\vec{g}_1(\cdot)$ and $\langle\vec{v}\rangle$ is the
  arithmetic mean of the components of vector $\vec{v}$.

  The second algorithm is described in Algorithm \ref{alg:vamp} and is known as vector AMP
  (\textit{VAMP}) \cite{vamp-rangan}. $\text{Diag}(\vec{v})$ is a diagonal matrix with elements of
  $\vec{v}$ on the diagonal. This algorithm has been shown to converge for a larger class of mixing
  matrices, namely those that are right-rotationally invariant\footnote{Unlike the sub-Gaussian
  assumption, given $\vec{A} = \vec{U}\text{Diag}(\vec{s})\vec{V}^T$, $\vec{s}$ does not follow a
  particular distribution and $\vec{U}$ can be any orthogonal matrix.}. Damping is also supported
  for further robustness. Comparing AMP with VAMP will demonstrate whether the advantages provided
  by the latter are relevant for the separation tasks evaluated here.

  \begin{algorithm}
      \caption{AMP with damping}
      \label{alg:amp}
      \begin{algorithmic}[1] 
        \Require Matrix $\vec{A} \in \mathbb{R}^{M \times N}$, observation vector $\vec{y} \in
            \mathbb{R}^M$, denoiser $\vec{g}_1$, noise precision $\gamma_\omega$ and damping
            factor $\theta$
        \Ensure Estimated solution $\vec{\widehat{x}}_t$
        \State \textbf{Initialisation:} $t = 0$, $\vec{s}_{-1} = 0$, $\vec{r}_0 \ge 0$, $\gamma_0 \ge 0$
        \Repeat
          \State $\vec{\widehat{x}}_t = \theta\vec{g}_1(\vec{r}_t,\gamma_t) + (1 - \theta)\vec{\widehat{x}}_{t-1}$
          \State $\tau^{p}_t = \frac{N}{M} \gamma^{-1}_t \langle\vec{g}'_1(\vec{r}_t,\gamma_t)\rangle$
          \State $\vec{s}_t = \theta(\gamma^{-1}_\omega + \tau^{p}_t)^{-1}(\vec{y} - \vec{A}\vec{\widehat{x}}_t + \tau^{p}_t\vec{s}_{t-1}) + (1 - \theta)\vec{s}_{t-1}$
          \State $\gamma_{t+1} = \theta(\gamma^{-1}_\omega + \tau^{p}_t)^{-1} + (1 - \theta)\gamma^{-1}_t$
          \State $\vec{r}_{t+1} = \vec{\widehat{x}}_t + \gamma^{-1}_{t+1}\vec{A}^T\vec{s}_t$
          \State $t = t + 1$
        \Until{Terminated}
      \end{algorithmic}
  \end{algorithm}

  \begin{algorithm}
      \caption{VAMP with damping}
      \label{alg:vamp}
      \begin{algorithmic}[1] 
        \Require Matrix $\vec{A} \in \mathbb{R}^{M \times N}$, observation vector $\vec{y} \in
            \mathbb{R}^M$, denoiser $\vec{g}_1$, noise precision $\gamma_\omega$ and damping
            factor $\theta$
        \Ensure Estimated solution $\vec{\widehat{x}}_t$
        \State \textbf{Initialisation:} $t = 0$, $\vec{r}_0 \ge 0$, $\gamma_0 \ge 0$
        \State Compute `economy' SVD: $\vec{A} = \vec{U}\text{Diag}(\vec{s})\vec{V}^T$
        \State $\widetilde{\vec{y}} = \text{Diag}(\vec{s})\vec{U}^T\vec{y}$
        \State $R = \text{rank}(\vec{A})$
        \Repeat
          \State $\vec{\widehat{x}}_t = \theta\vec{g}_1(\vec{r}_t,\gamma_t) + (1 -
              \theta)\vec{\widehat{x}}_{t-1}$
          \State $\alpha_t = \langle\vec{g}'_1(\vec{r}_t,\gamma_t)\rangle$
          \State $\vec{\widetilde{r}}_t = (\vec{\widehat{x}}_t - \alpha_t\vec{r}_t)/(1 - \alpha_t)$
          \State $\widetilde{\gamma}_t = \gamma_t(1 - \alpha_t)\alpha_t$
          \State $\vec{d}_t = \gamma_\omega\text{Diag}(\gamma_\omega\vec{s}^2 +
              \widetilde{\gamma}_t\vec{1})^{-1}\vec{s}^2$
          \State $\gamma_{t+1} = \theta\widetilde{\gamma}_tR\langle\vec{d}_t\rangle / (N -
              R\langle\vec{d}_t\rangle) + (1 - \theta)\gamma_t$
          \State $\vec{r}_{t+1} = \vec{\widetilde{r}}_t +
              \frac{N}{R}\vec{V}\text{Diag}(\vec{d}_t/\langle\vec{d}_t\rangle)(\widetilde{\vec{y}}
              - \vec{V}^T\vec{\widetilde{r}}_t)$
          \State $t = t + 1$
        \Until{Terminated}
      \end{algorithmic}
  \end{algorithm}

  \section{Methods}
  \label{section:methods}
  To carry out source separation, four stages can be identified \cite{sparse-bofill}: analysis,
  mixing matrix estimation, source reconstruction and synthesis. Analysis refers to transforming the
  mixture signal into the time-frequency domain, while synthesis is the inverse of this and is
  applied to the estimated sources. Mixing matrix estimation determines $\vec{A}$, and source
  reconstruction estimates the sources by solving \eqref{eq:inst}. Determining the mixing matrix is
  beyond the scope of this paper, and is not implemented because of availability of the ground truth
  mixing matrices for the experiments. Nonetheless, successful techniques already exist for
  instantaneous mixtures \cite{sparse-bofill, xu-2009}.

  \subsection{Time-Frequency Representation}
  Signals were transformed into the time-frequency domain with the short-time Fourier transform
  (STFT) \cite{stft-allen}, with frame size $L=1024$ and 70\% overlap. The window function was
  chosen to be the Hann window \cite{windows-harris}. As this gives a complex-valued signal of two
  variables, $X_i(n, k)$, a further transformation is necessary for it to apply to \eqref{eq:inst}.
  Noting that the time-domain signal is real, the STFT bins that are complex conjugates can be
  discarded. The remaining bins can then be separated into real and imaginary parts, giving $L$ real
  coefficients per frame. Finally, the frames can be concatenated along the dimension of the bins to
  give a signal of a single variable, $t=nL+k$. The inverse problem is then
  \begin{equation}
    \label{eq:inst_tf}
    \vec{Y}(t) = \vec{A}\vec{X}(t) + \vec{W}(t),
  \end{equation}
  where $\vec{X}(t)$ and $\vec{W}(t)$ are the time-frequency representations of the source and noise
  signals, respectively. The difference between \eqref{eq:inst} and \eqref{eq:inst_tf} is simply a
  change in notation to indicate that the latter is in the time-frequency domain.

  After the aforementioned transformation, each frame was also truncated to $L=720$ bins. The
  truncation was carried out by discarding the high-frequency bins of each frame, which is
  equivalent to low-pass filtering. This is an optional step, and may be detrimental for certain
  applications. However, the computational complexity of AMP/VAMP is quadratic with respect to the
  frame length because of the block-based approach taken, so even modest truncations can lead to
  large runtime improvements.

  \subsection{Source Reconstruction using AMP}
  The source reconstruction stage is the primary focus of this paper, and was implemented using the
  AMP algorithms. To do this, the \verb|GAMPmatlab| \cite{GAMPmatlab} implementation was used for
  AMP and VAMP. As explained in Section \ref{section:intro}, a typical source separation problem is
  very small in dimension, but AMP is only accurate for large systems. To solve this, a block-based
  approach \cite{xu-2009} was taken in which blocks of samples were considered together so that

  \begin{equation}
    \label{eq:inst_block}
    \underbrace{%
    \begin{bmatrix}
      Y_1(0)\\
      \vdots\\
      Y_1({T}-1)\\
      \vdots\\
      Y_{M}({T}-1)\\
    \end{bmatrix}
    }_{\hat{\vec{Y}}}=
    \underbrace{%
    \begin{bmatrix}
      \vec{\Lambda}_{11} & \cdots & \vec{\Lambda}_{1N}\\
      \vdots & \ddots & \vdots\\
      \vec{\Lambda}_{M1} & \cdots & \vec{\Lambda}_{MN}
    \end{bmatrix}
    }_{\hat{\vec{A}}}
    \underbrace{%
    \begin{bmatrix}
      X_1(0)\\
      \vdots\\
      X_1({T}-1)\\
      \vdots\\
      X_{N}({T}-1)\\
    \end{bmatrix}
    }_{\hat{\vec{X}}},
  \end{equation}
  where $T$ is the block size and $\vec{\Lambda}_{ij}=A_{ij}\vec{I}_{T}$ is a diagonal matrix with
  diagonal elements all equal to $A_{ij}$. The noise, $\hat{\vec{W}}$, has been omitted from
  \eqref{eq:inst_block} due to space limitation, though its form is the same as $\hat{\vec{X}}$ and
  $\hat{\vec{S}}$. The dimensions of the problem given in \eqref{eq:inst_block} are $\hat{M}=MT$ and
  $\hat{N}=NT$, which means they can be controlled by varying the block size, $T$.  Experimentally,
  it was found that $T=L$ is a good choice, and that significant deviations from this can negatively
  impact the performance -- even if $T$ is a multiple of $L$.

  Although \eqref{eq:inst_tf} and \eqref{eq:inst_block} are both instantaneous models, the meaning
  of sparsity has changed. In the former sequential-based approach, sparsity corresponds to the
  number of active sources for a given time-frequency point. This is a straight-forward concept
  because it is assumed that a small number of sources are active for a single point
  \cite{sparse-bofill, duet-jourjine, sparse-yilmaz}. In the block-based approach, it applies to
  sources from \textit{several} points; there is nothing to say these points should be considered
  separately.

  In any case, sparsity was enforced by setting the prior to be the Bernoulli-Gaussian (BG)
  distribution. This probability distribution is characterised by three parameters: the sparsity,
  $\rho$, the mean, $\mu$, and the variance, $\sigma^2$. Although values must be given initially,
  learning these parameters, as well as the noise precision, $\gamma_\omega$, is supported by
  \verb|GAMPmatlab| via expectation-maximisation (EM) \cite{em-bg-amp}.

  \section{Experiments}
  \label{section:results}
  In this section, the AMP algorithms are evaluated using the Signal Separation Evaluation Campaign
  (SiSec) datasets \cite{sisec2008, sisec2010} for underdetermined mixtures. More specifically, only
  the instantaneous stereo speech mixtures from the development datasets are assessed, giving a
  total of six mixtures containing either three sources or four sources. Each audio signal also
  includes the mixing matrix used to map the sources to the mixtures, so estimating the mixing
  matrix was not necessary. To objectively measure the performance, the PEASS toolkit \cite{peass}
  was utilised, which provides both non-perceptual and perceptual measures, given in decibels and as a
  score out of 100, respectively. For a description of the measures and their purpose, the reader is
  referred to \cite{bss-eval, peass}.

  Table \ref{table:params} lists the values of the parameters that were set when using
  \verb|GAMPmatlab|. The values for the BG prior parameters were determined through experimentation,
  and learning of the parameters was disabled. In fact, it was found that enabling learning for
  $\rho$ leads to poor source estimates. However, it may be beneficial for $\mu$ and $\sigma^2$. On
  the other hand, learning was enabled for the noise precision, $\gamma_\omega$. Its initial value
  was determined by relating it to the signal-to-noise ratio (SNR).
  \begin{equation}
    \label{eq:noise}
    \text{SNR} = 10\log{\frac{\hat{N}}{\hat{M}}\rho(\mu^2 + \sigma^2)\gamma_\omega}
  \end{equation}
  Rearranging \eqref{eq:noise} and setting the SNR as $-40$ dB gives the value in Table
  \ref{table:params}. \verb|maxIter| refers to the maximum number of iterations for the AMP
  algorithms. Both \verb|maxIter| and $\theta$ were evaluated over several values to examine how
  they affected the separation performance.

  \begin{table}[t]
  \centering
  \caption[Parameter values chosen for AMP and VAMP]{Parameter values chosen for AMP and VAMP.}
  \label{table:params}
  $\begin{tabular}{*{2}{c}}
  \toprule
  Parameter & Value(s) \\
  \midrule
  \verb|maxIter| & $\{5,10,...,50\}$ \\
  $\theta$ & $\{1,0.95,...,0.5\}$ \\
  $\mu$ & $0$ \\
  $\sigma^2$ & $5$ \\
  $\rho$ & $0.6$ \\
  $\gamma_\omega$ & $\frac{\hat{M}}{\hat{N}}\frac{10000}{\rho\sigma^2}$ \\
  \bottomrule
  \end{tabular}$
  \end{table}

  \subsection{Results}
  In Table \ref{table:results}, the averaged results are given for AMP and VAMP with and without
  damping. For these results, \verb|maxIter| was set to 30 for AMP and 10 for VAMP, and the damping
  factor was set to $0.75$ for AMP and $0.95$ for VAMP. The scores have been separated for mixtures
  containing three sources (SP3) and four sources (SP4). With the exception of VAMP-damped for SP3,
  the algorithms were successful at separating the sources to an extent, as indicated by the SIR and
  IPS measures. The SAR and APS measures are particularly high, which suggests that these algorithms
  are good at suppressing artefacts; AMP is especially good in this regard. SP4 performance is
  significantly worse in general, though this may be because the direction of arrival of the sources
  are closer.

\begin{figure}
    \centering
    \begin{subfigure}[t]{0.235\textwidth}
      \centering
      \includegraphics[width=\textwidth]{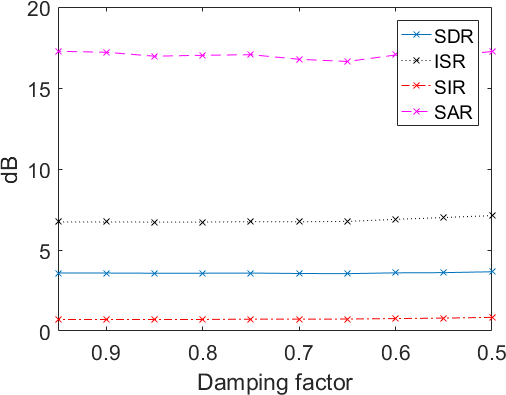}
      \caption{AMP/Non-perceptual}\label{fig:2a}
    \end{subfigure}
    \hfill
    \begin{subfigure}[t]{0.235\textwidth}
      \centering
      \includegraphics[width=\textwidth]{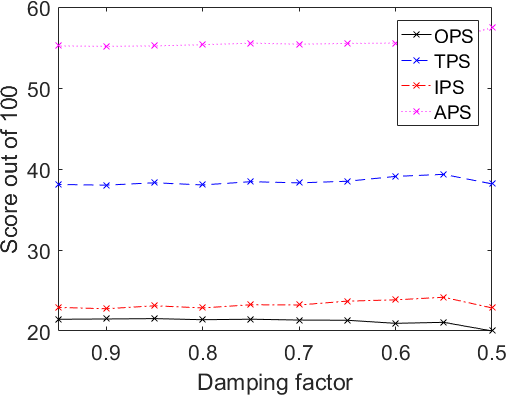}
      \caption{AMP/Perceptual}\label{fig:2b}
    \end{subfigure}
    \vskip\baselineskip
    \begin{subfigure}[t]{0.235\textwidth}
      \centering
      \includegraphics[width=\textwidth]{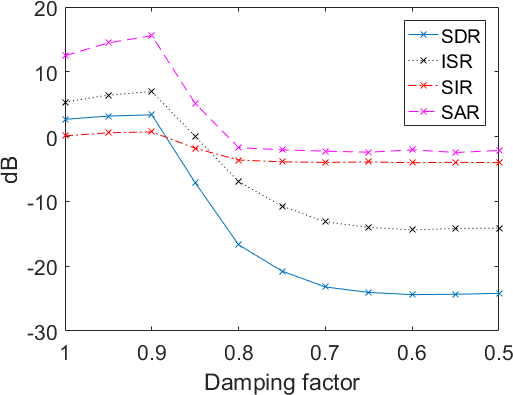}
      \caption{VAMP/Non-perceptual}\label{fig:2c}
    \end{subfigure}
    \hfill
    \begin{subfigure}[t]{0.235\textwidth}
      \centering
      \includegraphics[width=\textwidth]{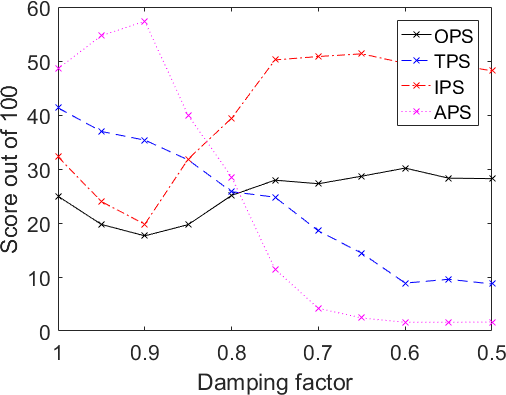}
      \caption{VAMP/Perceptual}\label{fig:2d}
    \end{subfigure}
    \caption[Performance of undamped AMP and VAMP as the damping factor varies]
            {Performance of undamped AMP and VAMP as the damping factor varies. Scores are averaged
             for mixtures with four sources. Damping does not appear to affect AMP, but it has a
             significant effect on VAMP, as the performance suddenly drops after $\theta \le 0.9$.}
    \label{fig:2}
  \end{figure}

  Comparing AMP with AMP-damped, there is very little difference between the two. To support this,
  Figures \ref{fig:2a} and \ref{fig:2b} plot how the performance of AMP varies with respect to
  $\theta$. It can be seen that the various measures hardly change as the amount of damping varies,
  suggesting that damping is unnecessary. On the other hand, even a small amount of damping has
  caused VAMP to fail to separate SP3 mixtures (as shown in Table \ref{table:results}). This is not
  the case for SP4 mixtures with $\theta=0.95$, but the plots in Figures \ref{fig:2c} and
  \ref{fig:2d} show that the same `phase transition" does in fact occur for SP4 but for $\theta \le
  0.9$. This suggests it depends on the number of sources.

\begin{figure}
    \centering
    \begin{subfigure}[t]{0.235\textwidth}
      \centering
      \includegraphics[width=\textwidth]{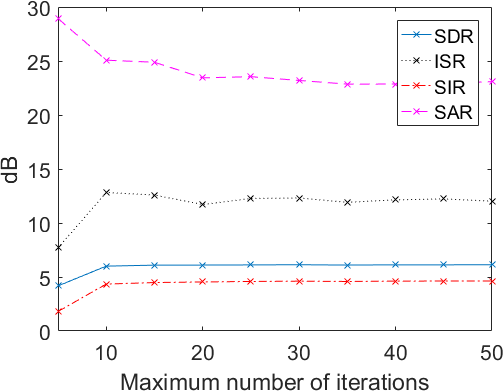}
      \caption{AMP/Non-perceptual}\label{fig:1a}
    \end{subfigure}
    \hfill
    \begin{subfigure}[t]{0.235\textwidth}
      \centering
      \includegraphics[width=\textwidth]{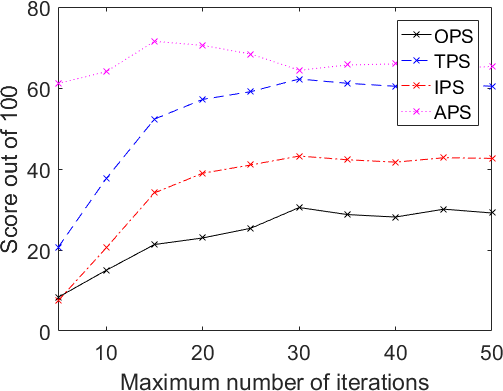}
      \caption{AMP/Perceptual}\label{fig:1b}
    \end{subfigure}
    \vskip\baselineskip
    \begin{subfigure}[t]{0.235\textwidth}
      \centering
      \includegraphics[width=\textwidth]{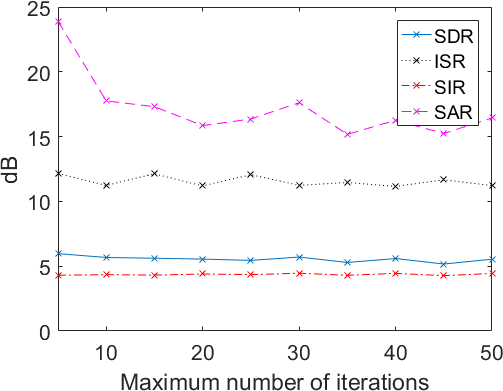}
      \caption{VAMP/Non-perceptual}\label{fig:1c}
    \end{subfigure}
    \hfill
    \begin{subfigure}[t]{0.235\textwidth}
      \centering
      \includegraphics[width=\textwidth]{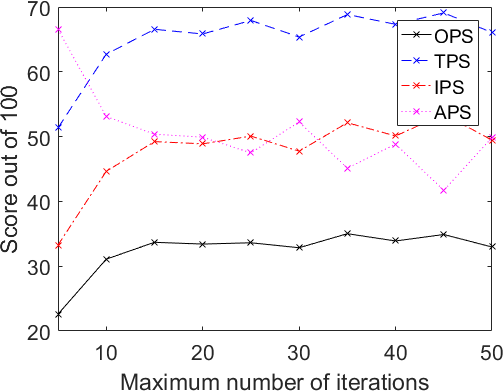}
      \caption{VAMP/Perceptual}\label{fig:1d}
    \end{subfigure}
    \caption[Performance of undamped AMP and VAMP as the number of iterations varies]
            {Performance of undamped AMP and VAMP as the maximum number of iterations varies. Scores
             are averaged for mixtures with three sources. The performance levels or converges after
             a certain number of iterations, with VAMP converging faster than AMP.}
    \label{fig:1}
  \end{figure}

  Finally, Figure \ref{fig:1} plots how the performance of undamped AMP and VAMP varies with respect
  to \verb|maxIter|. The scores tend to converge after a certain number of iterations, but in some
  cases with artefact suppression, the scores are actually higher for a low iteration count. This is
  because the algorithms have not separated the sources effectively at this point, so there is less
  chance of artefacts. Comparing the perceptual measures, we see that VAMP has converged faster.

  \begin{table}[t]
  \centering
  \caption[Averaged results for each algorithm using the ground truth mixing matrix]{Averaged results
  for each algorithm. The results have been separated in terms of the number of sources.}
  \label{table:results}
  \setlength{\tabcolsep}{4.5pt}
  \resizebox{\columnwidth}{!}{%
  $\begin{tabular}{*{9}{c}}
  \toprule
  Algorithm & \multicolumn{4}{c}{Speech (N=3)} & \multicolumn{4}{c}{Speech (N=4)}\ \\
      & SDR & ISR & SIR & SAR
      & SDR & ISR & SIR & SAR \\
      & OPS & TPS & IPS & APS
      & OPS & TPS & IPS & APS \\
  \midrule
  AMP & \textbf{6.1} & \textbf{12.3} & \textbf{4.6} & \textbf{23.2} & \textbf{3.6} & \textbf{6.7} &
  \textbf{0.7} & \textbf{17.3} \\
               & 30.5 & 62.2 & 43.2 & 64.4 & 21.4 & 38.1 & 22.9 & 55.2 \\[1ex]
  AMP-damped & \textbf{6.1} & \textbf{12.3} & \textbf{4.6} & 22.9 & \textbf{3.6} &
             \textbf{6.7} & \textbf{0.7} & 17.1 \\
           & 30.4 & 62.1 & 42.7 & \textbf{65.0} & 21.5 & 38.4 & 23.2 & \textbf{55.5} \\[1ex]
  VAMP & 5.7 & 11.2 & 4.3 & 17.8 & 2.7 & 5.3 & 0.1 & 12.5 \\
       & \textbf{31.1} & \textbf{62.7} & \textbf{44.6} & 53.1 & \textbf{25.0} & \textbf{41.3} &
  \textbf{32.3} & 48.6 \\[1ex]
  VAMP-damped & -10.0 & -0.3 & -1.1 & -2.0 & 3.2 & 6.4 & 0.6 & 14.5 \\
            & 23.9 & 56.5 & 68.1 & 1.9 & 19.7 & 36.9 & 24.0 & 54.7 \\
  \bottomrule
  \end{tabular}$
  }
  \end{table}

  \section{Conclusion}
  \label{section:conclusion}
  In this paper\footnote{This research was funded by the EPSRC project EP/K014307/2.}, approximate
  message passing algorithms were investigated for underdetermined audio source separation.
  These algorithms can be considered as approximations to belief propagation, offering a
  probabilistic framework for inferring the sources. Using a Bernoulli-Gaussian prior for the
  sources in the STFT (time-frequency) domain, we used AMP to solve the sparse linear inverse
  problem. Since AMP is only accurate for large systems, a block-based approach was taken instead of
  solving the problem for each sample individually.

  Two algorithms known as AMP and VAMP were evaluated with the SiSec dataset and the PEASS toolkit.
  The results showed that separation is indeed possible using this approach, with AMP performing
  better than VAMP. In terms of artefact suppression, AMP produced very promising results. This is
  a desirable property in applications such as hearing assistance, where it is less tolerable for
  such artefacts to be present. In the future, we would like to apply AMP to convolutive mixtures,
  for which the block-based approach easily lends itself. Another possible development is structured
  sparsity, e.g. grouping the real and imaginary coefficients.

  \vfill\pagebreak

  \bibliographystyle{IEEEbib}
  \bibliography{paper}
\end{document}